\RequirePackage{fix-cm}
\documentclass[smallextended]{svjour3}       % onecolumn (second format)
\smartqed  % flush right qed marks, e.g. at end of proof
\usepackage{amsmath}
\usepackage{graphicx}
\usepackage{lineno}
\usepackage{array}
\usepackage{longtable}
\usepackage{natbib}
\usepackage{rotating}
\usepackage{multirow}
\usepackage{hyperref}
 \journalname{Scientometrics}

\newcommand*\patchAmsMathEnvironmentForLineno[1]{%
\expandafter\let\csname old#1\expandafter\endcsname\csname #1\endcsname
\expandafter\let\csname oldend#1\expandafter\endcsname\csname end#1\endcsname
\renewenvironment{#1}%
{\linenomath\csname old#1\endcsname}%
{\csname oldend#1\endcsname\endlinenomath}}%
\newcommand*\patchBothAmsMathEnvironmentsForLineno[1]{%
\patchAmsMathEnvironmentForLineno{#1}%
\patchAmsMathEnvironmentForLineno{#1*}}%
\AtBeginDocument{%
\patchBothAmsMathEnvironmentsForLineno{equation}%
\patchBothAmsMathEnvironmentsForLineno{align}%
\patchBothAmsMathEnvironmentsForLineno{flalign}%
\patchBothAmsMathEnvironmentsForLineno{alignat}%
\patchBothAmsMathEnvironmentsForLineno{gather}%
\patchBothAmsMathEnvironmentsForLineno{multline}%
}

\begin{document}

\title{Analyzing the relationship between text features and grants productivity}%\thanks{Grants or other notes


\author{Jorge A. V. Tohalino        \and
        Laura V. C. Quispe \and Diego R. Amancio %etc.
}


\institute{ Jorge A. V. Tohalino \and Laura V. C. Quispe \and Diego R. Amancio \at
              Institute of Mathematics and Computer Science, \\
              Department of Computer Science,  University of S\~{a}o Paulo, \\
              S\~{a}o Carlos, S\~{a}o Paulo,  Brazil \\
              \email{diego@icmc.usp.br}           %  \\
}

\date{Received: DD Month YEAR / Accepted: DD Month YEAR}

\maketitle

\begin{abstract}
Predicting the output of research grants is of considerable relevance to research funding bodies, scientific entities and government agencies. In this study, we investigate whether text features extracted from projects title and abstracts  {are able to identify productive grants}. Our analysis was conducted in three distinct areas, namely Medicine, Dentistry and Veterinary Medicine. Topical and complexity text features were used to identify predictors of productivity. The results indicate that there is a statistically significant relationship between text features and {grants productivity}, however such a dependence is weak. A feature relevance analysis revealed that the abstract text length and metrics derived from lexical diversity are among the most discriminative features. We also found that the prediction accuracy has a dependence on the {considered project language} and that topical features are more discriminative than text complexity measurements. Our findings suggest that text features should be used in combination with other features to assist the identification of relevant research ideas.
\keywords{language analysis \and productivity \and  {grants productivity} \and text analysis %\and
%\newline
% \{Keywords should be in alphabetical order with the first letter of each keyword in upper case. No more than five keywords should be used.\}}
}
\end{abstract}

%{. This information could be used by authors to provide a more informed decision on the choice of journals to publish. The analysis of subfields impact might also be useful to understand the dynamics of journals visibility along time. A more refined visibility information in journals could also assist editors in identifying promising subfields or subfields that are no longer representative in terms of visibility.}

\section{Introduction}

Science of science has emerged, in the last few years, as the research area devoted to study the mechanisms underlying research and its related aspects~\citep{fortunato2018science}. This area has investigated a large number of important questions, including the evolution of science, and more specifically patterns of collaboration, citation and contribution among scientific entities~\citep{ding2011scientific}. Many studies have shed light on several important issues related to many processes involved in the creation and dissemination of scientific manuscripts. For example, studies on the behavior of paper citation networks not only have characterized these evolving networks, but also have developed models to predict their behavior~\citep{thelwall2018could,zeng2017science}. Many studies have also sought linguistic patterns in the scientific literature~\citep{mckeown2016predicting}. Similar studies have used paper metadata to analyze and understand the behavior of authors, including their collaboration/citation patterns and contributorship patterns~\citep{correa2017patterns,}.
Another important area in science of science concerns the studies devoted to make predictions in many scenarios~\citep{acuna2012predicting}. Those studies are important because they favor more informed decisions, thus improving the design of research policies.
While the focus of most investigations in science of science, especially those in the predictive area, use data from papers, in this paper we probe whether it is possible to make predictions regarding research output using {data extracted from research projects.}

Writing research proposals represents an important part of scientists' work. While proposals themselves are usually not intended to be published, they are equally relevant because they may ultimately decide whether novel ideas are going to be further developed and possibly disseminated. Deciding thus which proposals are going to be funded
{is of paramount importance for the advancement of science}.
Those decisions should be as fair as possible and, in many desired situations, they should be devoid of any personal bias other than the expected quality criteria.
In this sense, it becomes interesting if an automatic approach could \emph{assist} (but not \emph{replace}) {the traditional evaluation of research proposals before and after it is funded (at least in some criteria).}
While being less prone to personal bias, another advantage associated with automatic approaches is their ability to make decisions in a much short period of time, when compared to the traditional human classification.
Similar approaches have already been employed with success in other areas. For example, the quality and style of texts have been assessed using machine learning methods~\citep{antiqueira2007strong,silva2012word}. A pattern recognition approach applied in the context of grants assessment could also shed light on the understanding of which factors are associated with strong grants. This could be particularly useful for early career scholars, as many of them have received little or no feedback.
In the current study, we touch these points by probing whether information retrieved from projects can be used to {predict grants productivity.}

{
While many factors may affect the perceived quality of  projects~\citep{markowitz2019words,boyack2018toward}, in this study we focus on the analysis of textual features of funded projects. Our main objective is to analyze if we can {automatically} (i.e. using machine learning methods) predict grants productivity via linguistic patterns. We focused on two types of textual attributes, namely:}
\begin{enumerate}

    \item \emph{Topical features}: our hypothesis here consists in probing whether projects on specific subtopics are more likely to yield at least one publication. While certainly there are differences in publication/citation patterns across different fields~\citep{piro2013macro}, this type of analysis will investigate if combinations of words are associated with grants productivity.

    \item \emph{Complexity features}: {complexity features are affected by many psycholinguistic factors~\citep{graesser2004coh}. Examples of complexity features includes lexical diversity and function words counts~\citep{markowitz2019words,graesser2004coh}. Several works have shown that complexity measurements could influence the perceived quality of scientific works~\citep{markowitz2019words,letchford2015advantage,letchford2016advantage,wager2016declarative}. Our hypothesis regarding complexity is that similar complexity patterns could also appear in research projects.  In other words, if the results obtained from grants are disseminated using similar linguistic patterns, one could expect that the patterns could also influence the perception of the results/conclusions being analyzed, and thus influence grants productivity. For example, if grants abstracts are written in a clear, simple and more assertive way, one could expect that the same properties could appear in manuscripts submitted for publication and those patterns could affect acceptance decisions.}

    %predict if the results obtained from grants will be disseminated as well using a similar psycholinguistic pattern
    %
    %-- to characterize {research projects}.

\end{enumerate}
%
%We first analyzed the influence of topical features. We also used complexity measurements -- such as lexical diversity and words concreteness -- to characterize {research projects}. While the latter is intended to capture linguistic patterns that are topic independent, the former is used to investigate whether projects on specific topics are more likely to yield at least one publication.

Our analysis was conducted in a subset of {research grants} funded by the S\~ao Paulo Research Foundation (FAPESP-Brazil). We selected {research grants} in three areas comprising the largest number of projects funded by FAPESP. We considered projects in the following areas: Medicine, Dentistry and Veterinary Medicine.
{Here we classified if a grant yielded or not at least one publication. Therefore, in our classification system, two classes were considered, according to the number of observed publications: (i) zero publications; and (ii) one or more publications. While the term productivity might be allusive to distinguishing a few from many publications, hereafter productivity is used in the context of discriminating class (i) from (ii).
%A grant was considered successful if it yielded at least one publication.
}

%This paper aimed at addressing the following research questions: (i) is there any textual pattern relating text features and proposals success? (ii)
Several interesting results could be found in our analysis. By considering only a balanced version of the datasets, we found that there is a relationship between text features and grants productivity in all considered areas. However, we only found a weak correlation. When comparing complexity and topical features, the latter turned out to be more effective to predict {grants productivity}.
%
%A slight lower accuracy was found for the other areas. These results suggest that both topical and complexity measurements plays a relevant role in identifying successful proposals in the considered dataset.
%
We also found that the ability to predict {productivity} depends on the considered language (English or Portuguese). The accuracy was higher when analyzing texts written in researchers' native language (Portuguese). A feature importance analysis revealed that the measurements capturing lexical diversity of abstracts are relevant features for {identifying productive grants in all three considered datasets.}
Our analysis also revealed that the best classifiers for the adopted features were those based on Decision Trees. All in all, the adopted framework provides evidence that text features might be relevant in the identification of productive grants. We believe that text features could be combined with other features in future works to improve the discriminative rate of the classification systems.

This manuscript is organized as follows. In Section \ref{sec:rel}, we present related works on features used to predict the output of scientific papers and projects. In Section \ref{sec:dataset}, we describe the methodology used in the machine learning framework. The obtained results are discussed in Section \ref{sec:res}. Perspectives for works extending our approach are presented in Section \ref{sec:conc}.

\section{Related works} \label{sec:rel}

{Several studies have investigated the factors leading to the success of scientific items~\citep{markowitz2019words,wang2008measuring,boyack2018toward,xie2015modeling}.} In the case of scientific papers, many factors have found to play a role in defining their visibility. \cite{eom2011characterizing} show that the number of citations received in recent years can be an indication of future success. The authors proposed a linear preferential attachment with time dependent initial attractiveness that can recover not only the distribution of citations, but also the citation burstiness effect~\citep{eom2011characterizing}.  Similar models have extended this idea to characterize and predict researchers' impact. Other factors affecting the popularity of papers include the visibility of authors, journals, universities and the interdisciplinarity of fields and subfields~\citep{didegah2013factors,onodera2015factors,silva2016using}.

Text factors have also been found to affect the visibility of papers~\citep{amancio2012three,letchford2015advantage,paiva2012articles,mckeown2016predicting}. \cite{amancio2012three}  proposed a model to describe the evolution of papers citation networks. In addition to the age and visibility factor, they found that the similarity with other papers also represents a factor that cannot be disregarded.
%to improve the quality of the proposed model.
The impact of text features
%and more specifically the title
has also been discussed in some works~\citep{letchford2015advantage,paiva2012articles}. Recent results have pointed out that journals publishing papers with short titles tend to be more visible, as measured by the average citation counts. This is consistent with the idea that the use of a less complex linguistic style in papers leads to a better paper understanding. The influence of other textual factors on citations including question marks and titles describing results has also been reported~\citep{paiva2012articles}.

The factors affecting the success of research proposals have also been analyzed in the last few years~\citep{boyack2018toward,horlesberger2013concept,cabezas2013reviewers,fang2016nih,li2015big}. \cite{boyack2018toward} found that researchers productivity can not be used to predict proposals success. Likewise, institutional research strengths are not strong indicators of success. The success of research proposal was found to be more correlated with the topic similarity between the  proposal references and the respective applicant publications.

{\cite{markowitz2019words} studied if word patterns extracted from
National Science Foundation (NSF) proposals are able to predict the received amount of funding. As descriptors of text complexity the author used word counts, words per sentence, the percentage of common words and the complexity of thinking as measured via function words. Several interesting results were found showing a relationship between text variables and the amount of money received. Larger grant abstracts with fewer common words were among the main patterns correlating with funding success. \cite{markowitz2019words} advocated that the observed patterns contradicts NSF guidelines, since more complex textual structures are more correlated to funding success.}

Another feature that could be used to predict research proposal success are those related to peer review scores. In~\citep{cabezas2013reviewers}, the correlation between peers' scores and visibility indexes was analyzed for Spanish researchers in 23 fields. The study found that correlations are strongly dependent on the field being analyzed. Moreover, this study revealed that the main indicators that are associated to the acceptance of research proposals are the total number of publications and the number of papers published in prestigious journals. \cite{fang2016nih} studied the correlation between future research productivity and peers' scores of grants funded by the U.S. National Institutes of Health (NIH). They found that assigned scores are poor discriminators of success. As a consequence, they argue that this finding might increase the lack of discontentment with the peer review evaluation~\citep{germain2015healing}. Leading to a different conclusion, \cite{li2015big} argue that good peer review rating are correlated with better research outcomes, even when some specific controls are considered in the analysis, including authors and institutions visibility. This conclusion was reached in a dataset comprising $130,000$ research projects funded by NIH.
While many studies have focused on a variety of features to predict projects success, here we focus on text features to predict productivity. %and more specifically on the complexity/style language features.

\section{Material and Methods}

The dataset used in the current paper is described in Section \ref{sec:dataset}. The framework proposed to {classify grants} comprises the following three main steps:

\begin{enumerate}

    \item \emph{Feature extraction}: this phase is responsible for extracting topical and complexity features from textual fragments of {research projects}. This is detailed in Section \ref{sec:features}. While we test the influence of topical features, our main focus here is to analyze the influence of text complexity on the predictability of {grants productivity}.

    \item \emph{Pattern recognition}: the features extracted are used as input for traditional machine learning methods. An overview of methods is provided in Section \ref{sec:ml}. A more detailed reference on machine learning and pattern recognition methods can be found in~\citep{duda2012pattern}.

    \item \emph{Feature relevance analysis}: this phase is responsible for identifying the most relevant (i.e. discriminative) features. A brief description of the adopted method in provided in Section \ref{sec:gini}.

\end{enumerate}

\subsection{Dataset} \label{sec:dataset}

The main objective of this work is to analyze whether textual features can be used to predict the {productivity of research grants}. The adopted dataset consists in a subset of research projects carried out by researchers in Brazil (S\~ao Paulo State) and funded by the S\~ao Paulo Research Foundation\footnote{\href{www.fapesp.br/en}{fapesp.br/en}} (FAPESP). While it would be of interest to analyze the full content of research projects, this information is not public available. For this reason, most of the text analysis was based in two parts of the {research projects}: their title and abstract. The data were retrieved from the \emph{Biblioteca Virtual} website\footnote{\href{https://bv.fapesp.br/en/6/regular-grants-2-year-grants/}{bv.fapesp.br/en/6/regular-grants-2-year-grants}}.
{The association between papers and projects are automatically extracted by the \emph{Web of Science}\footnote{\href{www.webofknowledge.com}{webofknowledge.com}} dataset, and this information is also made available by the \emph{Biblioteca Virtual} website. We could not use information from other datasets since the link between paper and projects is not available. Once a project funded by FAPESP leads to a research paper, it is required that the authors acknowledge S\~ao Paulo Research Foundation. Specifically, the funding agency provides a format that every funded project must comply with. A grant must be mentioned using the explicit form ``aaaa/nnnnn-d'', where ``aaaa''
represents the year and  ``nnnnn-d'' is the grant number\footnote{This information is available from this \href{ https://fapesp.br/11789/referencia-ao-apoio-da-fapesp-em-todas-as-formas-de-divulgacao}{link} (in Portuguese).}.
%
%Therefore, papers that are not available in the Web of Science dataset are not considered.
%While this could be a major issue
}

The research projects are written originally in Portuguese. This is the reason why we focus our analysis on Portuguese textual data. However, because several abstracts are also available in English, we also provide an analysis of the dependence of the results on the considered language.

{We focused our analysis on regular grants, which are grants under the responsibility of a Principal Investigator associated with higher education (or research institutions) in the State of S\~ao Paulo.
Regular grants are usually limited to a duration of 2 years\footnote{Details regarding regular grants are available at \href{https://fapesp.br/apr}{fapesp.br/apr} (in Portuguese).}. Financial resources associated to each grant are usually limited to a maximum of $200,000$ \emph{Brazilian Reais}.}
We decided to analyze this type of grants for two main reasons: regular grants have a duration of at least 18 months (most of them lasts for 24 months). Therefore, some publications can be expected after this period. The other reason for choosing regular grants is the fact there are several grants of this specific type in the dataset. Considering this type of research project, we could retrieve textual information from  more than 31,000 instances. We considered projects funded between 1989 and 2015. More recent projects were disregarded because papers resulting from the projects may take several months to be published.

There are several useful bibliometric metrics to gauge research {grants productivity}. This could be the number of published papers, the number of citations, and several other metrics commonly used in quantifying success in academia~\citep{wang2013quantifying}. Because most of these distributions are skewed, we decided to simplify the criteria to consider a research project as productive.
%See e.g. Figure \ref{} showing the distribution of the number of papers and citations.
To avoid an extreme unbalancing in the number of positive and negative examples~\citep{li2010learning}, we consider a project as productive if it yielded at least one publication. While this criterion still generates unbalanced datasets, a considerable number of both positive and negative examples can be recovered.

{In order to avoid bias when comparing different research areas, we compared only projects belonging to the same area. In particular, we considered the following three areas comprising most of the research projects funded by FAPESP: Medicine (MED), Dentistry (DENT) and Veterinary Medicine (VET). \emph{According to the adopted criterion}, the percentage of positive examples in each area was: 41.27\%, 48.48\% and 31.96\% for Medicine, Dentistry and Veterinary Medicine, respectively. Additional basic statistics regarding grants productivity is available in Table~\ref{ttabsehaq}. Note that, in all cases, the number of positive examples is lower than the number of negative examples. In order to balance the data, the following standard procedure was applied~\citep{duda2012pattern}. Before training the models, we randomly draw from the set of negative examples $X$ instances, where $X$ is the number of positive instances in the dataset. This procedure was repeated 10 times for each area. The reported results therefore represents an average over these 10 generated balanced datasets.

\begin{table*}[h]
\centering
\caption{\label{ttabsehaq}Fraction of grants with a respective number of papers. In column \#P, $n+$ corresponds to $n$ or more papers being published. Note that, e.g. in the Medicine VET dataset, only a small fraction of grants published three or more papers ($5.5\%$).}
\begin{tabular}{c|c|c|c}
\hline \hline
\multirow{2}{*}{\textbf{\#P}} & \multicolumn{3}{c}{\textbf{Research areas}} \\ \cline{2-4}
 & {MED} & {DENT} & {VET} \\ \hline
2+ & 17.8\% & 25.6\% & 12.7\% \\
3+ & 9.3\% & 12.9\% & 5.5\% \\
4+ & 4.9\% & 7.3\% & 3.2\% \\
5+ & 3.1\% & 4.3\% & 1.5\% \\
6+ & 2.5\% & 3.3\% & 0.9\% \\
7+ & 1.6\% & 2.4\% & 0.6\% \\
8+ & 1.1\% & 1.3\% & 0.4\% \\
\hline \hline
\end{tabular}
\end{table*}

\subsection{Feature Extraction} \label{sec:features}

{Several studies have used linguistic features to analyze scientific items~\citep{larrimore2011peer,markowitz2014writing} We used two distinct approaches to analyze the abstract of the grants. Topical and complexity features. While topic features are intended to analyze if specific topics are associated with productive grants, complexity measurements analyzes whether language simplicity is associated with a lower or higher degree of productivity.}

In this paper, for each research project, we extracted textual features from both Portuguese and English text versions of research project abstract and titles. We are particularly interested in analyzing if there is an association between text structure (or complexity) and the observed research output. For comparison purposes, we also studied how predictable are {grants output} when texts are characterized with topical features.
%The following family of text features to characterize the abstracts.

The first feature used is the frequency of specific words. For each text, this generates a sparse vector whose $i$-th element stores the frequency of $i$-th word of the vocabulary. We also used a normalized version of this strategy, the so-called term frequency–inverse document frequency (tf-idf) approach. According to this strategy, the relevance of a word $w$ in each document depends not only on the frequency of $w$ in the document, but also on how many documents of the dataset. More specifically, the tf-idf representation of a word $w$ in a document (i.e research project) $d$ is given by:
\begin{equation}
    \textrm{tf-idf}(w,d) =  \frac{f(w,d)}{n_d} \cdot \frac{\log N}{\log{(N_{w})}},
\end{equation}
where $f(w,d)$ is the frequency of $w$ in $d$, $n_d$ is the number of words in $d$, $N$ is the number of documents in the dataset and $N_w$ is the number of documents in which $w$ occurs at least once.

{A different approach to characterize texts is via complexity analysis, that can be measure in different ways.
%
%\textcolor{blue}{DIZER NOSSA HIPOTESE AQUI COM RELACAO A COMPLEXIDADE}
%
A statistical approach based on structural features of text modeled as networks was described by~\cite{amancio2012complex}. While this approach was able to assess the readability of texts, the co-occurrence can only be effectively applied in larger texts~\citep{amancio2015probing}. The measurements used in the current are a subset of metrics adapted from the English version of Coh-Metrix~\citep{graesser2004coh}. }
{We used Coh-Metrix because it encompasses many different degrees of text complexity, including words, sentence, textbase and  situation model features~\citep{graesser2004coh}.}
Some examples of textual complexity features used here are:
\begin{enumerate}

    \item \emph{Basic counts}: total number of sentences, words, adjectives, adverbs and verbs. {Larger pieces of texts, such as word counts or more words per sentence, are associated with higher degree of complexity in texts~\citep{markowitz2019words,kincaid1975derivation}.}

    \item \emph{Logic operators}: this feature quantify the number of logical operators. {Note that logical operators such as ``if'' and ``or'' could denote texts with a certain degree of uncertainty~\citep{larrimore2011peer}. In the context of grants, a higher degree of uncertainty could mean taking more risks, and this could affect the productivity related to the grant.}

    \item \emph{Function word diversity}: this corresponds to the total of function word types (i.e. function word vocabulary size) normalized by the total number of different words (vocabulary size). {According to~\cite{pennebaker2014small}, particular function words reflect complex and analytic thinking. A style avoiding complex thinking could be associated with more clarity when expressing ideas. If such a simplicity is also reflected when writing papers, clarity could increasing the likelihood of a paper being published. A link between function words and concreteness is discussed by~\cite{larrimore2011peer}. In other words, counting function words is a different approach to capture the number of concrete words in texts.}

    \item \emph{Preposition diversity}: this corresponds to the same counting in \emph{function word diversity}, but applied to prepositions only. {The number of prepositions could be linked to with high academic performance in college and thus potentially could be useful to identify academic performance in research~\citep{pennebaker2014small}.}

     \item \emph{Punctuation diversity}: this corresponds to the same counting in \emph{function word diversity}, but applied to punctuation marks only. {Texts with long sentences and  few punctuation marks may suggest that they are harder to understand~\citep{graesser2004coh}. Sentences with a few pauses may require a higher cognitive effort to be processed.}

    \item \emph{Noun SD}: this corresponds to the standard deviation in the number of nouns per sentence. {Nouns can refer to concepts, and a text involving many different concepts could indicate a higher degree of complexity~\citep{graesser2004coh}.}

    \item \emph{Brunet index}: {this index quantifies the lexical diversity in the text. It is computed as $\beta = v^\alpha$, where $\alpha = n^{-0.165}$, $v$ is the vocabulary size and $n$ is the total number of words in the text. Typically, $10 \leq \beta \leq 20$.
    A high value of $\beta$ corresponds to a high lexical diversity. Thus higher values of $\beta$ indicate a richer language~\citep{brunet1978vocabulaire}.}

    \item \emph{Mean noun phrase}: {this corresponds to the average number of noun phrases in sentences. A noun phrase usually includes a noun and its modifiers. The interpretation of this measure in terms of complexity is similar to the one provided in \emph{Noun SD}. The difference is that here we are also considering modifiers along with nouns.}

    \item \emph{Concreteness SD}: this index quantifies the number of concrete words in the text. A concrete word is defined as a word representing  concepts and events that can be measured and observed. Examples of concrete words are `car' and `beans'. Conversely, examples of abstract words include `faith' and `chaos'.
    {
    As discussed by~\cite{larrimore2011peer}, the use of concrete words is related to a better contextualization of concepts and is linked to a reduction of uncertainty. Thus, more confident language could be linked to stronger results, which could make it easier for authors to publish papers.
    While issues with some semantic psycholinguistic variables have been reported~\cite{pollock2018statistical}, we decided to use this measurement because it has been useful in other contexts~\citep{diller2014effective}.
    A less context-dependent approach to operationalizing concreteness was used by~\cite{markowitz2019words}. The approach employed by~\cite{markowitz2019words} relies on function words, rather than a selection of concrete words. The method is consistent with function word features used in this work~\citep{graesser2004coh}.
    }

    \item \emph{NE ratio text}: this index corresponds to the proportion of named entities in the text. A named entity is any real-world entity, such as persons, locations, organizations, products etc~\citep{nadeau2007survey}. {In the scientific context, named entities could be linked e.g. to different methodologies or datasets. Texts with more different methodologies and/or named concepts involved could indicate a more detailed research, which in turn could facilitate publications from the respective grant. }

\end{enumerate}

The full list of the considered features and a detailed description of each feature can be found in~\citep{scarton2010coh}.

\subsection{Machine Learning Methods} \label{sec:ml}

%{TIRAR ACCURACY RATE DO ARTIGO E COLOCAR F1-SCORE (INCLUIR NA DESCRICAO??}

The textual features extracted from the abstract of the research projects are used in the classification process~\citep{duda2012pattern}. For each instance, we consider two possible classes: (i) zero publications; and (ii) one or more publications. In a typical classification task, the dataset is divided into two parts: the training and test datasets. The training dataset is used to create the model (e.g. a Decision Tree), while the test dataset is used evaluate the performance of the model. Here we used a standard procedure to split the original dataset into training and test datasets, the so called 10-fold cross validation scheme~\citep{duda2012pattern}.
The evaluation was based on the F1-Score measurement, a traditional measure in the area of information retrieval~\citep{manning2008introduction}.
To perform the classification the following algorithms were used:

\begin{enumerate}

    \item \emph{k-nearest neighbors} ($k$NN): in order to classify an unknown (unlabeled) instance, the algorithm first selects the $k$ nearest instances in the training dataset. The class associated to the unknown instance corresponds to the majority class observed in the selected $k$-set. The parameter $k$ is a parameter to be optimized~\citep{amancio2014systematic}. In the results section, we report the best results obtained for different values of $k$.

    \item \emph{Support Vector Machines (SVM)}: in this method, instances from different classes are divided by different spaces. These spaces are generated during the training phase. The main objective of this class of methods is to find a separation hyperplan between two or more classes. One of the main parameters of this methods is the kernel used to create the discriminative hyperplan. In this paper, we used the optimization strategy described in~\citep{amancio2014systematic}.

    \item \emph{Naive Bayes}: this method relies on the Bayesian optimal decision rule to perform a classification. Let $m = \{f_1,f_2,\ldots\}$ be the set of features used to characterize {research grants} (i.e., the features described in Section \ref{sec:features}).
    The class $\mathbf{c}$ assigned to a grant satisfies the following condition:
    \begin{equation}
        P(\mathbf{c}|m) \geq P(c_k|m),
    \end{equation}
    for every class $c_k \neq \mathbf{c}$, where $P(c_k|m)$ is the probability of the k-th class to have a set of features $m$. Because  $P(c_k|m)$ is not available in most cases, the Bayes' theorem can be used to find $\mathbf{c}$:
    \begin{equation}
        \mathbf{c} = \arg \max\limits_{c_k} \frac{P(m|c_k)} {P(m)} P(c_k).
    \end{equation}
    $P(m)$ is the same for every class $c_k$, therefore the above equation can be simplified to:
    \begin{equation}
        \mathbf{c} = \arg \max\limits_{c_k} P(m|c_k) P(c_k) = \arg \max\limits_{c_k} \big{[} \log P(m|c_k) + \log P(c_k) \big{]}.
    \end{equation}
    Assuming attribute independence, the class assigned to a new instance from the test dataset is computed as:
    \begin{equation}
        \mathbf{c} = \arg \max\limits_{c_k} \Bigg{[} \sum_{f_i} \log P(f_i|c_k) + \log P(c_k) \Bigg{]}.
    \end{equation}
    For the particular case of balanced datasets, $P(c_k)$ is uniform. Therefore,
    \begin{equation}
        \mathbf{c} = \arg \max P(m|c_k).
    \end{equation}

    \item \emph{Decision Trees}: the method based on Decision Trees uses a data structure composed of nodes and edges to represent the recognized patterns. In particular, a tree is a particular type of connected graph with the restriction that there is no cycle in such structure~\citep{cormen2009introduction}. Nodes represent attributes and edges correspond to the decision taken in different tests performed on the respective node. An example of decision tree is provided in Figure~\ref{fig:arvi}. The classification process starts at the root node (see Figure \ref{fig:arvi}) and continues until a leaf node (i.e. a node with no children) is reached. The class assigned to the instance in the test set     corresponds to the one stored in the respective leaf node. While this process is used to classify a new instance, a decision tree should be created during the training phase. This requires the definition of a measurement to identify the most discriminative attribute at each phase (i.e. node) of the classification process.
    A well-known measure used to identify the relevance of features is the Kullback–Leibler divergence. In the training dataset $D_\textrm{TR}$, the relevance of each feature $f_i$ is computed as:
    \begin{equation}
        \mathcal{K}(D_\textrm{TR},f_i) = \mathcal{H}(D_\textrm{TR}) - \mathcal{H}(D_\textrm{TR}|f_i).
    \end{equation}
    where $\mathcal{H}(D_\textrm{TR})$ is the entropy of the training dataset and $\mathcal{H}(D_\textrm{TR}|f_i)$ is the entropy of the training dataset considering the separation of classes obtained with the $i$-th feature~\citep{duda2012pattern,garreta2013learning}.

    \begin{figure}
    \centering
    \includegraphics[width=0.6\textwidth]{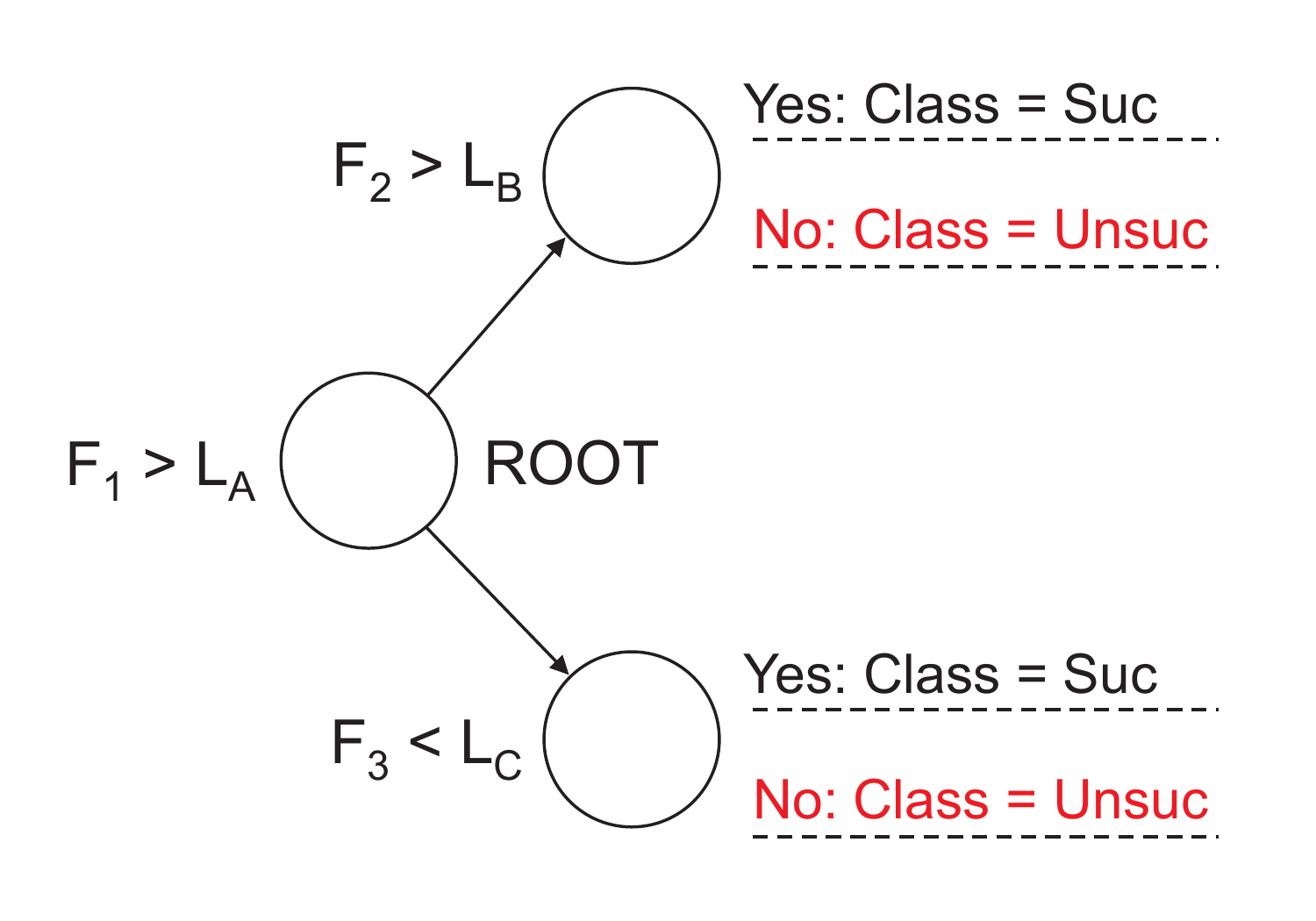}
    \caption{Example of decision tree used for classification. The classification process for a new instance starts at the root node. Consider a new instance that should be classified. This new instance is described by the the vector of features $(f_1 = x > L_A, f_2 = y < L_B, f_3 = z)$. The first test ($f_1>L_A$) leads the decision to the upper child node. Because the result of the current test fails (i.e. $f_2 > L_B$), this new instance is classified as yielding zero publications. In a similar fashion, an instance described by $(f_1 = q < L_A, f_2 = u, f_3 = v < L_C)$ would be classified as a {productive grant}.}
    \label{fig:arvi}
\end{figure}

    In addition to traditional decision trees, we also used random forests~\citep{breiman2001random}. The latter has the advantage of avoiding  the tendency of decision trees to overfit the training set~\citep{breiman2001random}. All results obtained with decision trees and random forests are reported as DTrees in the Results sections.

    \item \emph{Artificial Neural Networks (ANN)}: artificial neural networks are not a recent approach in the machine learning area, but have been widely used in recent years owing to the recent advancements in the deep learning area~\citep{lecun2015deep}. The most basic unit in a neural network is the perceptron. According to this model, the activation of a neuron depends on both input signals and transfers functions~\citep{hassoun1995fundamentals}. The activation can be considered as the perceptron output. Let $a_i$ be the $i$-th input and $w_i$ the weight associated to $a_i$. The output depends on the linear combination of input as weights, according to the value
    %
    %\begin{equation}
        $s = \sum_i w_i a_i + b$,
    %\end{equation}
    %
where $s$ is the input used as reference to the transfer learning function and $b$ is a constant value. The transfer learning function may assume many different forms~\citep{hassoun1995fundamentals}.
%
%The activation of the neuron will depend on transfer function $\phi(s)$, which defines the values of $s$ activating the considered neuron~\cite{hassoun1995fundamentals}.
%
If one chooses the Heaviside function, for example, the neuron if activated if $s$ is above an established threshold. The adequate choice of weights allows the neural network to effectively process the input in order to yield the expected output (class). Several algorithms have been designed to establish optimized synaptic weights~\citep{hassoun1995fundamentals}. One simple approach is to initially assign random weights and then update the values according to the observed error, i.e. the difference between the generated and expected outputs. Here we considered as neural network approach the multi-layer perceptron (MLP)~\citep{hassoun1995fundamentals}, a simple yet effective approach in many scenarios~\citep{amancio2014systematic}.
%
%the network to
%effectively process the input signals in order to generate the expected output.

\end{enumerate}

\subsection{Textual complexity measurements relevance} \label{sec:gini}

In order to evaluate the relevance of features when identifying {productive grants}, we used a feature relevance method that is based on decision trees. The relevance method uses the Gini impurity measurement to decide how discriminative is a partition of the dataset~\citep{nembrini2018revival}. The Gini impurity is defined as the probability of incorrectly classifying an instance if it were randomly classified according to the class distribution observed in the dataset. It is computed as:
\begin{equation} \label{eq:gini}
    \mathcal{G} = \sum_{i \in \mathcal{C}} p_i (1-p_i),
\end{equation}
where $\mathcal{C}$ is the set of classes. In our study, $\mathcal{C} =  \{\textrm{productive}, \textrm{zero publications}\}$. $p_i$ is the probability of choosing an instance from the $i$-th class in the considered subset.

As depicted in Figure~\ref{fig:arvi}, each tree node is associated with a feature. A feature is relevant in a node if it yields a decrease in the Gini impurity ($\Delta \mathcal{G}$) for the considered dataset. The decrease in impurity for each tree node is computed as
%other words, for each tree node we are given a $\Delta \mathcal{G}$, computed as
%
\begin{equation} \label{eq:ginivar}
    \Delta \mathcal{G} = \mathcal{G}_\textrm{B} - \beta_L \mathcal{G}_\textrm{L} - \beta_R \mathcal{G}_\textrm{R},
\end{equation}
where $\mathcal{G}_\textrm{B}$ is the Gini impurity before the dataset is split in the respective node and $\mathcal{G}_\textrm{L}$ and $\mathcal{G}_\textrm{R}$ are the Gini impurity obtained in the left and right child nodes, respectively. $\beta_L$ and $\beta_R$  are normalization factors to account for the number of instances falling in the left and right child nodes. This means that a higher weight is associated to the split region comprising more examples.
%
%Thus, the computation of $\Delta \mathcal{G}$ also considers how many examples leads to a left and right decrease in impurity.
%
%also takes into account the number of instances occurring in each side of the split characterized by the node under analysis.
%
Finally, the relevance of a given feature $m_i$ is computed as the average decrease in impurity observed in all nodes in which $m_i$ is used.

To illustrate the process of computing the Gini impurity for a given split of the dataset, we provide an example in Figure \ref{fig:gini}. The original dataset with two classes and two features is shown in the left panel.  Because there are 16 positive and 16 negative examples, the probability of misclassifying a randomly selected instance is $50\%$ (i.e. $\mathcal{G}_B = 50\%$). After the dataset is split (see right panel), two subsets are created. In the left subset, the impurity is zero, because all instances belong to the same class. In the right subset, the impurity is computed according to equation \ref{eq:gini}:
\begin{equation}
    \mathcal{G}_\textrm{R} = \frac{1}{17} \Bigg{(} 1 - \frac{1}{17} \Bigg{)} + \frac{16}{17} \Bigg{(} 1 - \frac{16}{17} \Bigg{)} = 0.1107.
\end{equation}
The proportion of data in the left and right subsets are respectively 15/32 and 17/32. Thus, the decrease in impurity, $\Delta \mathcal{G}$, as defined in equation \ref{eq:ginivar}, is given by:
\begin{equation}
    \Delta \mathcal{G} = \mathcal{G}_B - \frac{15}{32} \mathcal{G}_L - \frac{17}{32} \mathcal{G}_R = 0.4412.
\end{equation}
In other words, the split for the considered feature yield a reduction of $\Delta \mathcal{G} = 0.4412$ in the impurity of the original dataset.

\begin{figure}
    \centering
    \includegraphics[width=0.8\textwidth]{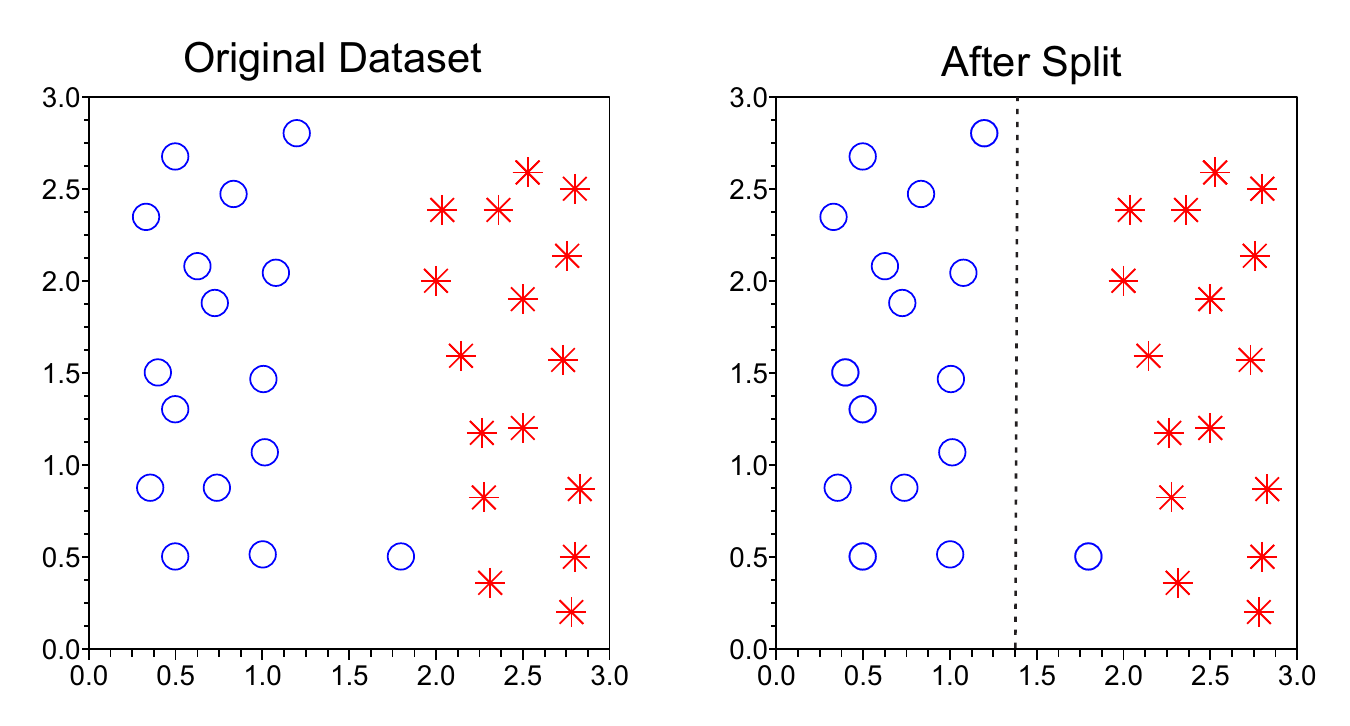}
    \caption{Computing the decrease in Gini impurity for a small dataset with two classes. For each class, there are $16$ instances. In the original dataset, the probability of misclassification is high for a randomly drawn instance is high, i.e. $\mathcal{G} = 0.50$. After the original dataset is split in two subsets, the discrimination of classes becomes almost perfect. This leads to a high variation in the Gini impurity, i.e. $\Delta \mathcal{G} = 0.4412$.  }
    \label{fig:gini}
\end{figure}

%{Adicionar uma figura ilustrando??}

\section{Results and Discussion} \label{sec:res}

In this section, we discuss the obtained results. Our analysis is divided into three sections. In Section \ref{sec:perf}, the  performance for different features and machine learning methods is reported. In Section \ref{sec:language}, we discuss whether the discriminability varies significantly when considering different languages (Portuguese and English). Finally, in Section \ref{sec:rel}, we perform an analysis of features relevance.

%Resultados gerais
% - Resultados cohmetrix
% - Resultados tf idf
% - Comparacao de complexity vs. topicality
% - Comparacao de topicality sem stopwrods (????)

%Language dependency
% - complexity and tf-idf

%Analise de features da complexity

\subsection{Performance analysis} \label{sec:perf}

In this section, we start the discussion of results by considering the F1-Score obtained with complexity measurements extracted from title and abstracts (in Portuguese). The obtained results are shown in Table \ref{tab:tabVesp}. We show, for each considered dataset (Medicine, Dentistry and Veterinary Medicine) the performance obtained from the machine learning methods considered in this study.
{We computed the statistical significance of the obtained results. If $n_g$ is the number of corrected classified grants, the $p$-value corresponds to the probability of correctly classifying at least $n_g$ instances just by chance. To compute the $p$-value we took into account that classes are imbalanced. This means that each single instance, in a random classification, is correctly classified with probability $p_i$, where $p_i$ is the fraction of the dominant class in the considered dataset.}
The best results for each dataset were found to be statistically significant. They are highlighted in Table \ref{tab:tabVesp}.
The best result in predicting the {productivity of a research grant} (according to the adopted productivity criteria) was found in the area of Veterinary Medicine. In this case a $\textrm{F1-Score} = 0.6008$ has been found.
Significant results were also found for Medicine and Dentistry, however with a lower discriminability result. In the best scenario, the F1-Score obtained for these area were $0.5673$ and $0.5725$, respectively.
These results suggest that the complexity of texts plays a statistically significant role in predicting research grants productivity. While the best results are significant, they reveal only a weak discriminative power.
%This could be an indication that
%could be a component that can be used to predict the scientific output of research proposals.
%
\begin{table*}[h]
\centering
\caption{F1-Score rate obtained when classifying grants productivity using Coh-metrix features~\citep{graesser2004coh} for Portuguese. Three different datasets were considered:  Medicine (MED), Dentistry (DENT) and Veterinary Medicine (VET). The best results for each dataset are highlighted. We also show for each F1-Score the corresponding significance of the classification. The best results were obtained with decision trees.
}
\begin{tabular}{l|c|c|c}
\hline
\hline
\multirow{2}{*}{\bf Method} & {Medicine} & {Dentistry} & {Vet. Med.} \\
    &  F1-Score & F1-Score & F1-Score \\
\hline
DTrees   &  ${\bf 0.5673}$ & ${\bf 0.5725}$ & ${\bf 0.6008 }$ \\
SVM   &  $0.5417$  & $0.5437 $ & $0.5657 $ \\
kNN   &  $0.4997$  & $0.5384 $ & $0.5336 $ \\
Bayes  &  $0.4182$ & $0.4800 $ & $0.5115 $ \\
MLP   &  $0.5177$  & $0.5142 $ & $0.5278 $ \\
\hline
\hline
\end{tabular}
\label{tab:tabVesp}
\end{table*}

When considering different strategies for supervised classification and the set of Coh-metrix features, in all three considered datasets, the best results were obtained with Decision Trees. The second best classifier was SVM in all considered datasets.
The worst classification systems, were found to be kNN and Naive Bayes. Interestingly, the performance of the Naive Bayes was much lower than the F1-Score obtained with decision trees, and such a difference turned out to be more prominent in the Medicine area.  This result suggests that the choice of classification systems plays an important role in the performance of the considered classification task, even when the same family of features are considered.

While the results shown in Table \ref{tab:tabVesp} considers as features only complexity factors of language, it would be interesting to analyze if improved results can be obtained when topical features are used to predict grants productivity. For this analysis, we considered the tf-idf representation of texts. Here, the classification considered different parts of the {research project}. In the experiments, we considered the title, the subject, a combination of title and subject, and two variations of the tf-idf representation of the abstracts. The subject is provided by researchers and corresponds to a few words representing the corresponding research area. For the tf-idf representation of abstracts, the adopted approach selected the $X$ most frequent words as features. In the approaches referred to as Abstract$^{(1)}$ and Abstract$^{(2)}$, we used $X = 1,100$ and $X = 7,196$ words, respectively.

In Table \ref{tab:tfidf}, we show the results obtained for different classifiers. The best results found for each dataset are  significant. The best F1-Score were $0.5766$, $0.6246$ and $0.6395$ respectively for the MED, DENT and VET datasets. The discriminability observed in the Veterinary Medicine area once again was found to be slightly higher than the discriminability found in other areas. These results also suggest that the frequency-based features also play a role in predicting productivity of the considered datasets.  Topical features were found to be more discriminative than complexity measurements, when considering the best results. The highest improvement in performance was found in the DENT dataset: the F1-Score improved from $0.5725$ (with complexity measurements) to $0.6246$ (with topic features).
%This means that for this area, topical information is more relevant than complexity measurements to predict the success of research grants.
%
Interestingly, the best performance was obtained with Decision Trees: in all three datasets, the highest accuracy was obtained with this classifier. %The multilayer perceptron also yielded results similar to those obtained with decision trees. %Differently from the results obtained with text complexity features, the SVM classifier achieved a low classification performance.
%
%title (591+- feats), Subject (S) (409+- feats), T + S (908+-), Abstract (1) (1100+- feats), Abstract (2) (7196+- feats)
%
\begin{table*}[h]
%\begin{adjustbox}{angle=90}
\centering
\caption{Results based on the frequency (tf-idf) considering different fragments of research projects: the title, the subject, a combination of title and subject and the abstract. For the latter strategy, we selected the $X$ most frequent words as features. In the approaches referred to as Abstract$^{(1)}$ and Abstract$^{(2)}$, we used $X = 1,100$ and $X = 7,196$ words, respectively. The best significant results for each dataset are highlighted.}
\begin{tabular}{l|c|c|c|c|c}
\hline
\hline
\multicolumn{1}{c|}{\multirow{3}{*}{\bf Features}} & \multicolumn{5}{c}{Research Projects on \emph{Medicine}} \\
    &  \multicolumn{1}{c}{DTrees} & \multicolumn{1}{c}{SVM}  & \multicolumn{1}{c}{kNN} & \multicolumn{1}{c}{Bayes} & \multicolumn{1}{c}{MLP} \\
        &  \multicolumn{1}{c}{F1-Score} & \multicolumn{1}{c}{F1-Score}  & \multicolumn{1}{c}{F1-Score} & \multicolumn{1}{c}{F1-Score} & \multicolumn{1}{c}{F1-Score} \\
\hline
Title	    & $0.5189 $ & $0.4972 $ & $0.4773 $ & $0.5228 $ & $0.5115 $ \\
Subject	    & $0.5376$ & $0.4349$ & $0.5163$ & $0.5372 $ & $0.5320$ \\
Tit. + Sub. & $0.5404$ & $0.4979$ & $0.4978$ & $0.5254$ & $0.5158$ \\
Abstract$^{(1)}$ & ${\bf 0.5660}$ & ${\bf 0.5695}$ & $0.5293$ & ${\bf 0.5588}$ & $0.5470$ \\
Abstract$^{(2)}$  & ${\bf 0.5766}$ & ${\bf 0.5649}$ & $0.5314$ & ${\bf 0.5696}$ & $0.5505$ \\
\hline
\multicolumn{1}{c|}{\multirow{2}{*}{\bf Features}} & \multicolumn{5}{c}{Research Projects on \emph{Dentistry}} \\
    &  \multicolumn{1}{c}{DTrees} & \multicolumn{1}{c}{SVM}  & \multicolumn{1}{c}{kNN} & \multicolumn{1}{c}{Bayes} & \multicolumn{1}{c}{MLP} \\
\hline
Title	    & $0.5541$ & $0.5055$ & $0.4755$ & $0.5606$ & $0.5636$ \\
Subject	    & $0.5664$ & $0.4680$ & $0.5466$ & $0.5636$ & $0.5646$ \\
Tit. + Sub. & ${\bf 0.5916}$ & $0.5133$ & $0.5284$ & ${\bf 0.5772}$ & ${ \bf 0.5771}$ \\
Abstract$^{(1)}$ & ${\bf 0.5979}$ & ${\bf 0.6033}$ & $0.5502$ & $0.5697$ & ${\bf 0.5913}$ \\
Abstract$^{(2)}$ & ${\bf 0.6246}$ & ${\bf 0.5770}$ & $0.5548$ & ${\bf 0.6026}$ & ${\bf 0.5869}$ \\
\hline
\multicolumn{1}{c|}{\multirow{2}{*}{\bf Features}} & \multicolumn{5}{c}{Research Projects on \emph{Veterinary Medicine}} \\
    &  \multicolumn{1}{c}{DTrees} & \multicolumn{1}{c}{SVM}  & \multicolumn{1}{c}{kNN} & \multicolumn{1}{c}{Bayes} & \multicolumn{1}{c}{MLP} \\
\hline
Title	    & $0.5485$ & $0.5115$ & $0.5127$ & $0.5543$ & $0.5235$ \\
Subject	    & $0.5566$ & $0.4885$ & $0.508$ & $0.5517$ & $0.5496$ \\
Tit. + Sub. & $0.5721$ & $0.5503$ & $0.5128$ & $0.5579$ & $0.5407$ \\
Abstract$^{(1)}$ & $0.5890$ & $0.5903$ & $0.5371$ & $0.5799$ & $0.5636$ \\
Abstract$^{(2)}$ & ${\bf 0.6395}$ & $0.5886$ & $0.5076$ & $0.5958$ & ${\bf 0.6010}$ \\
\hline
\hline
\end{tabular}
\label{tab:tfidf}
\end{table*}
%\end{adjustbox}

Table \ref{tab:tfidf} also reveals that particular fragments of {research projects} are more discriminative than others. Considering the best classifier in all three datasets (i.e. the decision tree method), we found that the best results occur when the abstract is taken into account.
In general, when more features are considered (i.e. Abstract$^{(2)}$), a higher discriminability rate is obtained. A lower performance is observed when considering both the title and the {grant subject}. However, the performance of Abstract$^{(1)}$ and  Tit. + Sub. is similar for Dentistry and Veterinary Medicine.
It is also worth noting that a very low discriminative power obtained with kNN in all three datasets. Regardless of the chosen set of features, the discriminability is always too low. This is consistent with the results observed for complexity measurements.
%
%improvement in performance occurs for the Naive Bayes when additional features are considered in the strategy based on the abstract. In all three datasets, we found an improvement of roughly 10\%.

\subsection{Language dependence} \label{sec:language}

As mentioned in Section \ref{sec:dataset}, the abstract of each {research project} is available in two languages: Portuguese and English. The results reported in Section \ref{sec:perf} were obtained for textual data in Portuguese. Here we analyze whether there is a significant difference in performance when considering abstracts in English.

The results obtained when considering complexity measurements are shown in Table \ref{tab:engcomp}. The best results for each language and research area are highlighted. When comparing the best results for Portuguese and English, we found no a difference in performance, especially in both Dentistry and Veterinary Medicine areas. In all three datasets, the best discriminability was found for the Portuguese language.
%A slightly difference in performance was found for the DENT dataset.
In the Veterinary Medicine area, the best discriminability rate found for Portuguese is roughly 13\% higher than the best score obtained using abstracts in English.
%We also note that good results for English are obtained with Decision Trees, and MLP.

\begin{table*}[h]
\centering
\caption{F1-Score obtained when discriminating {research projects} productivity. We used complexity features to characterize the texts. The results reveal that a difference in performance is observed when comparing Portuguese and English abstracts. The best significant results for each dataset and language are highlighted.}
\begin{tabular}{l|c|c}
\hline
\hline
\multicolumn{1}{c|}{\multirow{3}{*}{\bf Method}} & \multicolumn{2}{c}{Projects on \emph{Medicine}} \\
    &  \multicolumn{1}{c}{Portuguese} & \multicolumn{1}{c}{English}\\
        &  \multicolumn{1}{c}{F1-Score} & \multicolumn{1}{c}{F1-Score}  \\
\hline
DTrees	& ${\bf 0.5673 } $ &  ${0.5322} $  \\
SVM     & $0.5417 $ &  $0.4855  $  \\
kNN     & $0.4997  $ &  $0.5069  $  \\
Bayes   & $0.4182  $ &  $0.4876  $  \\
MLP   & $0.5177  $ &  $0.5180 $  \\
\hline
\multicolumn{1}{c|}{\multirow{2}{*}{\bf Method}} & \multicolumn{2}{c}{Projects on \emph{Dentistry}} \\
    &  \multicolumn{1}{c}{Portuguese} & \multicolumn{1}{c}{English}\\
\hline
DTrees	& ${\bf 0.5725} $ &  ${0.5104} $  \\
SVM     & $0.5437$ &  $0.4858$  \\
kNN     & $0.5384$ &  $0.5087$  \\
Bayes   & $0.4800$ &  $0.5048$  \\
MLP   & $0.5142$ &  $0.4953$  \\
\hline
\multicolumn{1}{c|}{\multirow{2}{*}{\bf Method}} & \multicolumn{2}{c}{Projects on \emph{Vet. Med.}} \\
    &  \multicolumn{1}{c}{Portuguese} & \multicolumn{1}{c}{English}\\
\hline
DTrees	& ${\bf 0.6008} $ &  $0.5168$  \\
SVM     & $0.5657$ &  $0.4843$  \\
kNN     & $0.5336$ &  $0.5038$  \\
Bayes   & $0.5115$ &  $0.4948$  \\
MLP   & $0.5278$ &  ${0.5308} $  \\
\hline
\hline
\end{tabular}
\label{tab:engcomp}
\end{table*}

In Table \ref{tab:tfeng}, we show the results obtained for the English datasets when considering tf-idf features. The best results for each dataset and language are also highlighted. The analysis of the best results reveals a dependence with language that is similar to the one observed in Table \ref{tab:engcomp}: in all three datasets, we found a difference in performance when comparing the best results obtained for abstracts written in Portuguese and English.
The best results for English occur with Decision Trees.

%The multilayer perceptron classifier also displayed a good performance in abstracts written in English.
%
\begin{table*}[h]
\centering
\caption{F1-Score obtained when discriminating {research project} productivity. We used tf-idf features to characterize the {project} abstracts. The best results for each dataset and language are highlighted.
%
%Accuracy rate obtained when discriminating research proposals as successful or unsuccessful. We used tf-idf to classify the texts.
%The results reveal that only a small difference in performance is observed when comparing Portuguese and English abstracts.
}
\begin{tabular}{l|c|c}
\hline
\hline
\multicolumn{1}{c|}{\multirow{3}{*}{\bf Method}} & \multicolumn{2}{c}{Projects on \emph{Medicine}} \\
    &  \multicolumn{1}{c}{Portuguese} & \multicolumn{1}{c}{English}\\
        &  \multicolumn{1}{c}{F1-Score} & \multicolumn{1}{c}{F1-Score}  \\
\hline
DTrees	&${\bf 0.5766}$ & $0.5488$     \\
SVM     &${\bf 0.5695}$ & $ {0.5555}$     \\
kNN     &$0.5314$ & $0.5406$     \\
Bayes   &${\bf 0.5696}$ & $0.5354$    \\
MLP   &  $0.5505$ & $0.5155$    \\
\hline
\multicolumn{1}{c|}{\multirow{2}{*}{\bf Method}} & \multicolumn{2}{c}{Projects on \emph{Dentistry}} \\
    &  \multicolumn{1}{c}{Portuguese} & \multicolumn{1}{c}{English}\\
\hline
DTrees	& ${\bf 0.6246}$ & $0.5164$     \\
SVM     & ${\bf 0.6033}$ & $0.4934$     \\
kNN     &  $0.5548$ & $0.4979$    \\
Bayes   & ${\bf 0.6026}$ & $0.5543$     \\
MLP   & ${\bf 0.5869}$ & ${0.5413}$    \\
\hline
\multicolumn{1}{c|}{\multirow{2}{*}{\bf Method}} & \multicolumn{2}{c}{Projects on \emph{Vet. Med.}} \\
    &  \multicolumn{1}{c}{Portuguese} & \multicolumn{1}{c}{English}\\
\hline
DTrees	& ${\bf 0.6395}$ & $0.5231$     \\
SVM     & $0.5903$ &  $0.4932$   \\
kNN     & $0.5371$ & $0.5299$    \\
Bayes   & $0.5958$ & $0.5509$     \\
MLP     & ${\bf 0.6010}$ & ${0.5653}$     \\
\hline
\hline
\end{tabular}
\label{tab:tfeng}
\end{table*}

We note that abstracts in English are less discriminative, and this  might be related to the fact that almost all {research projects} are written by Portuguese native speakers. Because English can be viewed as a second language in the analyzed {project}, a lower discriminability might be a consequence of a lower linguistic variety, both at the complexity and topical levels. In other words, textual properties observed in {projects} might  have a higher variability when the text is written in the researcher's native language. As a consequence, this effect can cause significant differences in the considered classification task. This result suggests that predicting productivity with text features should also take into account whether the language being analyzed is a first or second language.

%The accuracy rates shown in Tables \ref{tab:engcomp} and \ref{tab:tfeng} suggest that the discriminability of research proposals depends upon the language, at least for the considered research areas and languages. This suggests that a machine learning approach applied to other languages is a tool to identify proposals more likely to yield at least one publication. Such an approach could consider both complexity and topical features, as no significant different have been found when comparing both types of features.

\subsection{Feature relevance} \label{sec:rel}

The results in the previous section showed that there is a dependence between text features extracted from {research projects} and the output of the respective grants. The productivity of specific grants according to tf-idf features might be a consequence of the fact that some subjects and topics are more visible than others, for several reasons~\citep{mckeown2016predicting,silva2016using}. A similar behavior has been reported at the journal level, since interdisciplinary papers tend to accrue more citations than papers that are specific to a single discipline~\citep{leydesdorff2019interdisciplinarity,leydesdorff2011indicators}. The importance of text complexity (i.e. topic independent) features is not as clear.  In order to better understand in future works if particular features plays a more relevant role in predicting the productivity of grants, in this section we provide an analysis of the main complexity features responsible for identifying productive grants.

For the analysis of features relevance, we used the strategy described in Section \ref{sec:perf}, which is based on the Decision Tree algorithm. We used this strategy because Decision Trees displayed excellent results in the previous performance analysis. For each dataset, we ranked in decreasing order the complexity features according to the value of $\Delta \mathcal{G}$, which corresponds to the average decrease in impurity for tree nodes involving that feature.
Because of the cross-validation and balancing procedures, the ranking obtained by each feature varies in each considered subset of the dataset. In Figure \ref{fig:rank} we show the ranking diagram depicting the average rank of the best ranked features for each research area.
\begin{figure}
    \centering
    \includegraphics[width=0.95\textwidth]{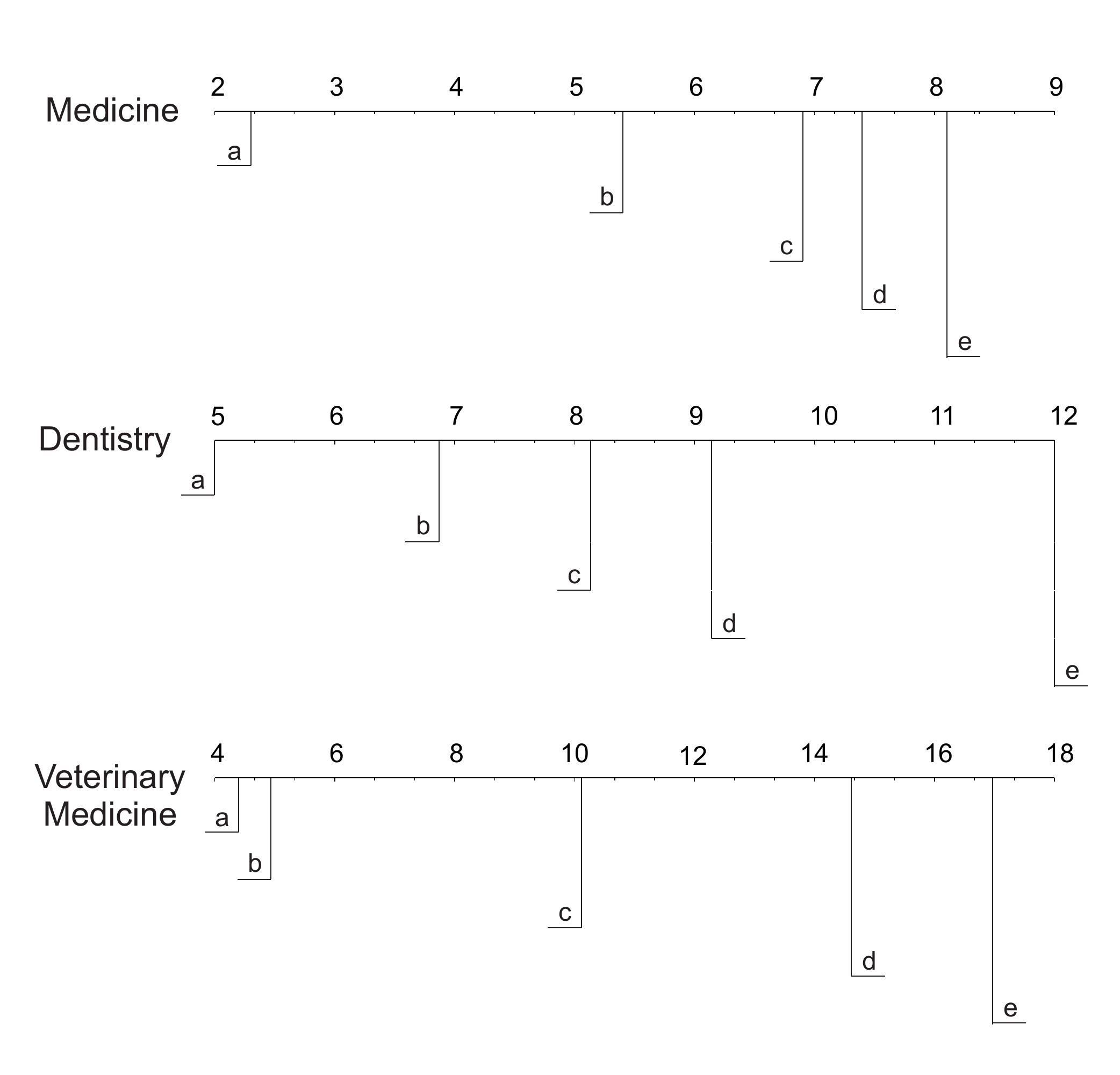}
    \caption{Feature ranking diagram for the classification of research projects {according to their productivity}. For each dataset, we show the average ranking obtained by each of the considered Coh-metrix features. In Medicine, the best features were: (a) function word diversity; (b) nouns SD; (c) total number of words; (d) preposition diversity; and (e) Brunet index. In Dentistry, the best features were (a) mean noun phrase; (b) total number of words; (c) preposition diversity; (d) punctuation diversity; and (e) concreteness SD. In Veterinary Medicine, the best features were (a) NE ratio text; (b) total number of words; (c) noun ratio; (d) brunet index; and (e) preposition diversity.}
    \label{fig:rank}
\end{figure}

An analysis of Figure \ref{fig:rank} revealed that the best ranked features (in decreasing order) for each of the considered datasets were:
\begin{enumerate}

    \item \emph{Medicine}: (a) function word diversity, (b) standard deviation of noun occurrences, (c) total number of words, (d) preposition diversity and (e) Brunet index.

    \item \emph{Dentistry}: (a) mean noun phrase, (b) total number of words, (c) preposition diversity, (d) punctuation diversity and (e) standard deviation of words concreteness.

    \item \emph{Veterinary Medicine}:  (a) named entity ratio text, (b) total number of words, (c) noun ratio, (d) Brunet index and (e) preposition diversity.

\end{enumerate}
While some features, in average, seems to be considerably better than others  in the diagram, the Critical Difference~\citep{demvsar2006statistical} (not shown in the diagram) reveals that there is no significant difference among these 5 best ranked features.

Some interesting patterns can be observed from the best ranked features. First, the total number of words seems to be a relevant feature for the classification. {However, it is not possible to identify a single pattern (e.g. a correlation) between this feature and {grants} output, since this feature can be used in different ways in different tree nodes}. Other features that were found to be relevant for the classification accuracy are the Brunet index and the preposition diversity. These measurements show that not only the abstract length is important, but also the diversity of lexical items. This finding is compatible with studies correlating lexical diversity and writing quality~\citep{antiqueira2007strong}.
{The relevance of preposition diversity reveals that not only the diversity of semantic concepts might be relevant to discriminate {productive grants}, but also stopwords (prepositions), i.e. words conveying no semantical meaning. This result reinforces the importance of style when performing a text analysis in grants~\citep{markowitz2014writing,markowitz2019words}. Because the presence of function words could indicate a lack of concrete words (see e.g.~\cite{larrimore2011peer}), we can infer that concrete words are also a relevant feature for the task. Therefore, it seems that concrete words are not only important to detect the amount of funding~\citep{markowitz2019words}, but also if the grant will yield a paper.}
In addition, the 'concreteness' of words -- as quantified by Coh-Metrix -- also seems to play a role in the identification of productive DENT grants. This means that concreteness might be an important feature less dependent of the considered research area.
%Such a relevance, though, is not evident in the other datasets, meaning that some features might be relevant only in particular research areas.

All in all, the results obtained in this section showed that particular word choices and the ability to construct a rich vocabulary might be correlated with the output observed in  {research grants}. From a linguistic point of view, it should be interesting to investigate in future works if any of the identified relevant features (and respective patterns) can be considered as marks of a high-quality writing. {If papers resulting from well-written projects are themselves written in a similar high-quality style, one should expect that they are more likely to be published (provided that all other paper requirements and standards are met). This could explain the fact that the above features are relevant to detect {productive grants}}.

\section{Conclusion} \label{sec:conc}

The development and advancement of science is fundamental for the evolution of society. A driving force towards the development of science are the preliminary ideas, which often lead to important developments in the near (or distant) future. While many ideas should be developed without restriction, in practice a limitation in resources hinders all research ideas from being developed at their highest potential. In practical terms, this means that many research proposals are not funded, and this may affect the success and diffusion of important ideas. In this context, it is clear that funding decisions should be as effective as possible in order to avoid the waste of resources that could be otherwise invested in truly \emph{strong} ideas.

Despite some criticisms, the role of peer review in identifying promising ideas remains undeniable~\citep{kassirer1994peer}. As it happens in other bibliometrics contexts, it is still interesting to provide automatized tools that can assist humans in particular issues~\citep{silva2016using,amancio2015comparing,daud2015using}. In this context, in this paper, we analyzed whether textual features extracted from research projects can be used to identify productive grants. Given the nature of our dataset, we considered a machine learning setting where productive {research grants} were those yielding at least one publication. As features, we focused on two types of linguistic attributes. First, we used complexity measures that are topic-independent. We also used, for comparison purposes, a simple frequency-based approach. A dataset of research grants funded by S\~ao Paulo Research Foundation (FAPESP-Brazil) was considered and analyzed in three distinct areas, namely Medicine, Dentistry and Veterinary Medicine.

Our analysis revealed several interesting findings. We found that there is indeed a relationship between text features and grants productivity. However, the use of text features alone showed only a weak discriminability. Interestingly, the subject being approached (i.e. topical features) seems to be more relevant than the style (i.e. complexity) of the text provided in the title and abstract of the respective project.
We also found that the obtained results are prone  to the considered language, since similar differences in performance were found for projects written in English and Portuguese. A feature relevance analysis also revealed that text length and the vocabulary diversity are among the most discriminative features.

%First, we found a high accuracy when using complexity measurements to characterize research proposals abstracts. We found an accuracy of 83.3\% in a binary classification with decision trees in research proposals in the area of Dentistry.
%Similar results have been found for the other studied areas (Medicine and Veterinary Medicine).

%This result was found to be as good as the one obtained when classifying texts with tf-idf.

%Considering complexity measurements, we also found that among the evaluated classifiers, excellent performance was obtained with Decision Tree and SVM for all three considered datasets.

The results of this paper suggest that both complexity and topical features have a potential to identify productive {research grants}, according to the adopted criteria for productivity. However, there is a large space for improvement in performance, since other features can be used to characterize {projects}.
%We believe that text analysis has a potential to \emph{assist} the analysis of research proposals.
{In this paper, we limited the sense of productivity by considering that productive grants are those yielding at least one publication.} In future works, it is interesting to analyze other productivity criteria, including e.g. the total number of publications, the reputation of the respective journals and conferences and other measurements derived from citation and usage counts~\citep{ruan2020predicting,hou2020social}. We also intend to incorporate additional features in order to improve our predictions, including text network-based attributes~\citep{stella2020forma,stella2019forma,amancio2015concentric,stella2019modelling} and other features related to researchers and their respective institutes~\citep{correa2018word,de2016using}.

\begin{acknowledgements}
D.R.A. acknowledges CNPq-Brazil (grant no. 304026/2018-2) for financial support. This study was financed in part by the Coordenação de Aperfeiçoamento de Pessoal de Nível Superior - Brasil (CAPES) - Finance Code 001.
\end{acknowledgements}

\newpage

\bibliographystyle{spbasic_updated}      % basic style, author-year citations
%\bibliography{references}   % name your BibTeX data base

\end{document}